\documentclass[Journal,twocolumn,10pt]{IEEEtran}

\usepackage{graphicx}
\usepackage{amsfonts}
\usepackage{amssymb}
\usepackage{amsmath}
\usepackage{fancyhdr}
\usepackage{lastpage}
\usepackage{subfigure}

\newcommand{\comment}[1]{}

     \newcommand{\qed}{\nobreak \ifvmode \relax \else
           \ifdim\lastskip<1.5em \hskip-\lastskip
           \hskip1.5em plus0em minus0.5em \fi \nobreak
           \vrule height0.75em width0.5em depth0.25em\fi}
\begin{document}


\title{Reduced Complexity Demodulation and Equalization Scheme for Differential
Impulse Radio UWB Systems with ISI }

\author{\authorblockN{Xudong Ma \\}
\authorblockA{FlexRealm Silicon Inc., Virginia, U.S.A.\\
Email: xma@ieee.org} }

\maketitle

\begin{abstract}

In this paper, we consider the demodulation and equalization problem
of differential Impulse Radio (IR) Ultra-WideBand (UWB) Systems with
Inter-Symbol-Interference (ISI). The differential IR UWB systems
have been extensively discussed recently \cite{choi02},
\cite{feng05}, \cite{franz03}, \cite{ho02}, \cite{hoctor02},
\cite{souilmi03}, \cite{witrisal}. The advantage of differential IR
UWB systems include simple receiver frontend structure. One
challenge in the demodulation and equalization of such systems with
ISI is that the systems have a rather complex model. The input and
output signals of the systems follow a second-order Volterra model
\cite{witrisal}. Furthermore, the noise at the output is data
dependent. In this paper, we propose a reduced-complexity joint
demodulation and equalization algorithm. The algorithm is based on
reformulating the nearest neighborhood decoding problem into a mixed
quadratic programming and utilizing a semi-definite relaxation. The
numerical results show that the proposed demodulation and
equalization algorithm has low computational complexity, and at the
same time, has almost the same error probability performance
compared with the maximal likelihood decoding algorithm.

\end{abstract}

\section{Introduction}

\label{section_introduction}

Ultra-WideBand (UWB) communication systems have attracted much
attention recently. The UWB communications have the advantages of
robustness due to multi-path diversity,  low possibilities of
intercept and high location estimation accuracy. UWB systems are
favorable choices for short range high bit rate communications or
medium-to-long range low bit rate communications. For example, UWB
systems have been considered for video communications in Wireless
Personal Area Networks (WPAN). In this case, the transmission rates
can be as high as 400M bits per second. UWB communication systems
have also been considered for Wireless Sensor Networks (WSN) as a
low-power and low-cost solution. The FCC (US Federal Communications
Commission) has recently approved the use of UWB communications and
allocated a spectrum range  of $7.5$ GHz for UWB communications.

A communication system is considered to be a UWB system, if the
system's bandwidth spans more than $1.5$ GHz, or $25\%$ of the
center frequency. The UWB systems transmit data by sending pulses,
each with very small time duration. For one transmitted pulse, a
large number of replicas of the same pulse are received at the
receiver side due to multi-path. The number of resolvable
multi-paths can be as high as more than $100$ as shown in
\cite{foerster02}. As a consequence, multi-path diversities are
automatically achieved. However, accurate channel estimation can be
quite complex and difficult.

The existing approaches for UWB communications include,
Direct-Sequence (DS) UWB, Multi-Band (MB) UWB, and low-complexity
non-coherent Impulse Radio (IR) UWB systems. The DS-UWB systems use
direct sequence spreading technique to convert information signals
into wide-band signals, \cite{win00}, \cite{eshima}. Under the
condition that the channel estimation is accurate, the RAKE receiver
is the optimal demodulation scheme. However, the channel estimation
for UWB channels is difficult and complex. Without the information
about the correct RAKE weights, the systems suffer a performance
loss by using sub-optimal RAKE structures (for example, equal weight
combining).

MB-UWB systems are recently proposed and discussed in
\cite{saberinia03}, \cite{foerster03}, \cite{batra04}. The MB-UWB
systems use the Orthogonal Frequency Division Multiplexing (OFDM)
technology. The advantage of MB-UWB systems include higher
achievable bit rates, flexibility in spectrum occupation, good
coexistence with narrow band communications. The disadvantages
include complex architectures, and high power consumption. The third
class of UWB systems is  the non-coherent IR UWB systems. In such
systems, complete channel estimation is not required. Therefore, the
channel estimation constraint is greatly relaxed.

In this paper, we consider a low-complexity non-coherent IR UWB
system - the differential IR UWB system proposed in \cite{ho02}. In
the differential IR UWB systems, the transmitted information is
differentially encoded. At the receiver side, a low-complexity
Autocorrelation (AcR) receiver is adopted. The decoding decision
variables are autocorrelations
\begin{align}
\int_{t_0}^{t_1}r(t)r(t+\delta) dt,
\end{align}
where, $r(t)$ is the received signal, and $\delta$ is the time
difference between two consecutive pulses. The integral can be
implemented either in the analog domain or in the digital domain. In
both cases, the decoder architecture is largely simplified.

One problem of the AcR receiver is that the transmitted messages and
the receiver decoding decision variables follow a nonlinear
second-order Volterra model, especially when
Inter-Symbol-Interference (ISI) is present in the systems
\cite{witrisal}. The maximal-likelihood sequential decoders can be
adopted, however their computational complexities generally grow
exponentially with the length of delay spread.

In this paper, we propose a  reduced-complexity demodulation and
equalization algorithm. The algorithm is based on a reformulation of
the nearest neighborhood decoding problem into a mixed quadratic
programming and a Semi-Definite Programming (SDP) relaxation. The
computational complexity of the proposed algorithm grows only
polynomially with respect to the block length and is independent of
the length of delay spread. We show by simulation results that the
performance loss caused by the proposed sub-optimal demodulation
algorithm is negligible.

SDP relaxation has been previously adopted to solve decoding
problems and combinatorial optimization problems. In \cite{goemans},
an approximation algorithm for maximum cut problem based on SDP
relaxation has been proposed. Detection algorithms for MIMO channels
based on SDP relaxation have also been proposed in \cite{nekuii},
\cite{sidiropoulos}, \cite{mobasher}, \cite{mobasher08},
\cite{wiesel05}. For interested readers, a review of SDP
optimization can be found in \cite{todd01}.

The rest of this paper is organized as follows. In Section
\ref{section_system}, we describe the system model. We present the
proposed demodulation and equalization algorithm in Section
\ref{section_decoding_algorithm}. Numerical results are presented in
Section \ref{sec_numerical}. Conclusions are presented in Section
\ref{sec_conclusion}.

Notation: We use the symbol $\pmb{\mathcal S}$ to denote the set of
symmetric matrices. Matrices are denoted by upper bold face letters
and column vectors are denoted by lower bold face letters. We use
$\pmb{A}\succeq 0$ to denote that the matrix $\pmb{A}$ is positive
semi-definite. The symbol $\otimes$ is used to denote the Kronecker
product. We use $\pmb{A}_{i,j}$ to denote the element of the matrix
$\pmb{A}$ at the $i$-th row and $j$-th column. We use $\pmb{a}_{i}$
to denote the $i$-th element of the vector $\pmb{a}$. We use
$\pmb{A}^T$ and $\pmb{a}^T$ to denote the transpose of the matrix
$\pmb{A}$ and the vector $\pmb{a}$ respectively. We use
$tr(\pmb{A})$ to denote the trace of the matrix $\pmb{A}$. The
function $\mbox{sign}(\cdot)$ is defined as,
\begin{align}
\mbox{sign}(x)=\left\{
\begin{array}{ll}
1, &  \mbox{if }x\geq 0 \\
-1, & \mbox{otherwise}
\end{array}
\right.
\end{align}

\section{System Model}
\label{section_system}

We assume that the message is transmitted in a block by block
fashion. The transmitted signal in one block is
\begin{align}
s(t)=\sum_{n=0}^{N_b-1}\sum_{i=0}^{N_p-1}a_i[n]\bar{w}\left(t-t_i[n]\right)
\end{align}
where $\bar{w}(t)$ is the transmitted pulse, $a_i[n]$ is the pulse
polarity for the $i$-th pulse of the $n$-th symbol, $t_i[n]$ is the
pulse time for the $i$-th pulse of the $n$-th symbol. Each block has
$N_b$ symbols, and each symbol corresponds to $N_p$ pulses.

Denote the data symbol by $d[n]\in\{-1,+1\}$. The data symbols are
differentially encoded as,
\begin{align}
a_i[n]=\left\{
\begin{array}{ll}
a_{N_p-1}[n-1]d[n-1]b_{N_p-1}, & \mbox{ if }i=0 \\
a_{i-1}[n]d[n]b_{i-1}, & \mbox{ otherwise}
\end{array}
\right.
\end{align}
where, $b_0,b_1,\ldots,b_{N_p-1}$ is the pseudo-random amplitude
code sequence, $b_i\in\{-1,+1\}$. The pulse time
\begin{align}
t_i[n]=nT_s+c_i
\end{align}
where $T_s$ is the symbol duration, $c_i$ is the relative pulse
timing. The relative pulse timeing $c_i$ is related to the
pseudo-random delay hopping code $\{D_i\}$,
\begin{align}
D_i=\left\{
\begin{array}{ll}
T_s+c_0-c_{N_p-1}, & \mbox{ if }i=N_p-1 \\
c_{i+1}-c_i, & \mbox{ otherwise}
\end{array}
\right.
\end{align}
The pseudo-random amplitude code and delay hopping code are used to
facilitate multiple access.

\begin{figure}[h]
 \centering
 \includegraphics[width=3in]{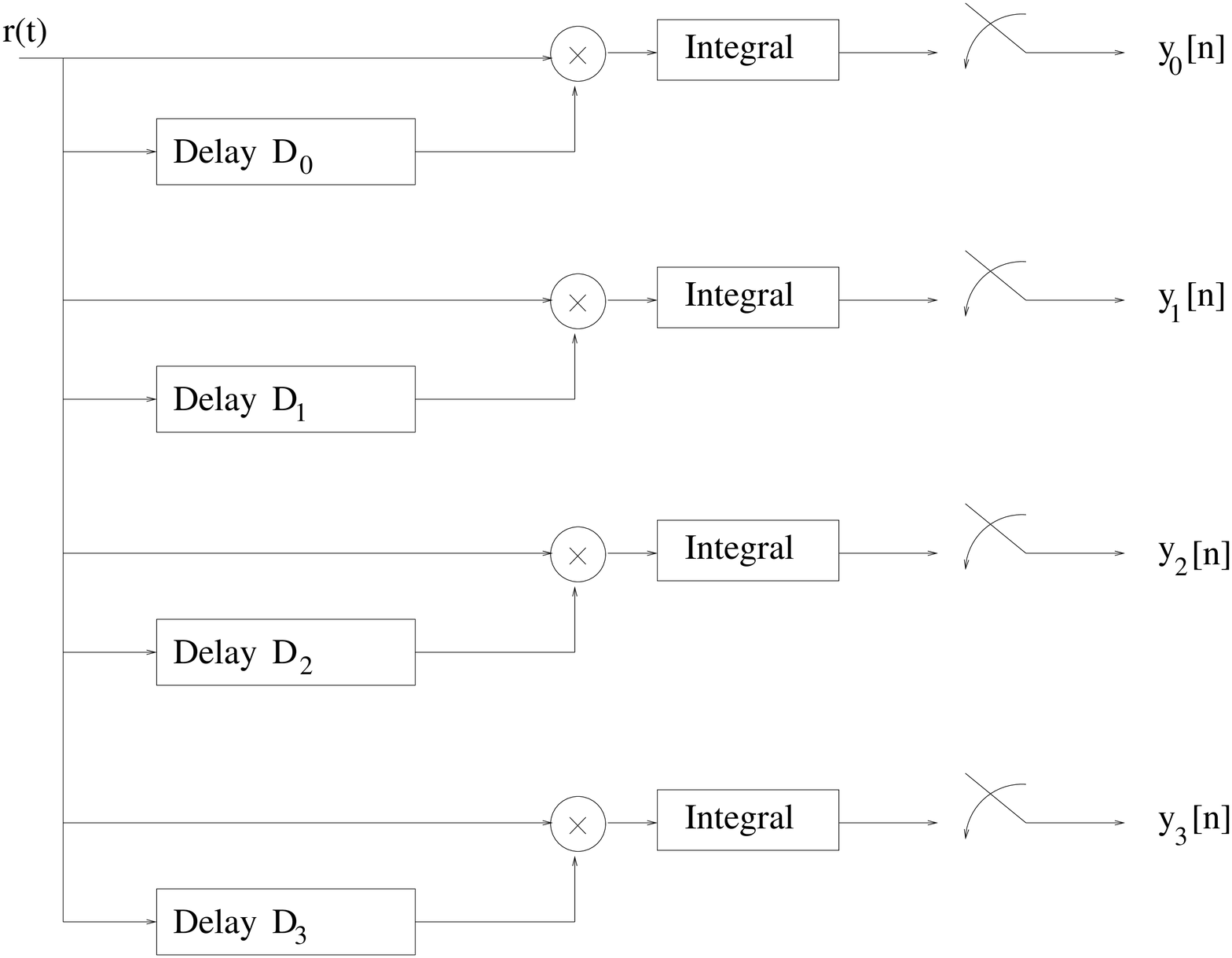}
 \caption{Block diagram of the autocorrelator receiver}
 \label{fig_auto_correlator_receiver}
\end{figure}

The received signal is,
\begin{align}
r(t)=\sum_{n=0}^{N_b-1}\sum_{i=0}^{N_p-1}a_i[n]g\left(t-t_i[n]\right)+n(t),
\end{align}
where, $g(t)$ is the channel response for the pulse $\bar{w}(t)$,
$n(t)$ is the noise. The receiver front end is shown in Fig.
\ref{fig_auto_correlator_receiver}. Denote the decoding decision
variable for the $n$-th symbol by $z[n]$,
\begin{align}
y_i[n]=\int_{t_i[n]}^{t_i[n]+T_I}r(t)r(t+D_i)dt,
\end{align}
\begin{align}
z[n]=\sum_{i=0}^{N_p-1}y_i[n]b_i,
\end{align}
where, $T_I$ is the integral time.

Let us define
\begin{align}
I_g(t_1,t_2;\tau)=\int_{t_1}^{t_2}g(t)g(t+\tau)dt
\end{align}
Denote the data vector by $\pmb{d}$,
\begin{align}
\pmb{d}=[d_0,d_1,\ldots,d_{N_b-1}]^T.
\end{align} Define the
column vector,
\begin{align}
\pmb{a}=\left[a_0[0],a_1[0],\ldots,a_0[n],a_1[n],\ldots,a_{N_p-1}[N_b-1]\right]^T.
\end{align}
Neglecting noise, we have
\begin{align}
y_i[n]=\pmb{a}^T\pmb{A}_i[n]\pmb{a}.
\end{align}
In the above equation, $\pmb{A}_i[n]$ is a matrix, such that the
$(n'N_p+i'+1,n''N_p+i''+1)$ element is
\begin{align}
I_g\left(t_{i}[n]-t_{i'}[n'],t_i[n]-t_{i'}[n']+T_I;t_{i'}[n']-t_{i''}[n'']+D_i\right).
\end{align}
Finally, the decoding decision variables can be written as,
\begin{align}
z[n]=\sum_{i=0}^{N_p-1}b_i\pmb{a}^T\pmb{A}_i[n]\pmb{a}=\pmb{a}^T\pmb{B}[n]\pmb{a},
\end{align}
where, $\pmb{B}[n]=\sum_{i=0}^{N_p-1}b_i\pmb{A}_i[n]$.

The vector $\pmb{a}$ can be written as,
\begin{align}
\pmb{a}=\pmb{Q}\left(\pmb{r}+\pmb{P}\pmb{d}\right).
\end{align}
In the above equation, $\pmb{Q}$ is a diagonal matrix,
\begin{align}
\pmb{Q}=\mbox{diag}[1,b_0,b_0b_1,\ldots],
\end{align}
\begin{align}
[\pmb{Q}]_{k,k}=\prod_{j=0}^{k-2}b_{j\mbox{mod}N_p}.
\end{align}
The matrix $\pmb{P}$,
\begin{align}
\pmb{P}=\pmb{I}_{N_b}\otimes \pmb{s},
\end{align}
where, $\pmb{I}_{N_b}$ is an identical matrix, and $\pmb{s}$ is a
vector with length $N_p$ of alternating $0$, $1$,
\begin{align}
\pmb{s}=[0,1,0,1,\ldots,1]^T.
\end{align}
The vector $\pmb{r}$ is,
\begin{align}
\pmb{r}=\pmb{i}_{N_b}\otimes (\pmb{i}_{N_p}-\pmb{s}),
\end{align}
where $\pmb{i}_{N_b}$ and $\pmb{i}_{N_p}$ are the all one column
vectors with length $N_b$ and $N_p$ respectively.

With the above notation, the second-order Volterra model of the
system is,
\begin{align}
z[n]=(\pmb{r}+\pmb{P}\pmb{d})^T\pmb{Q}^T\pmb{B}[n]\pmb{Q}(\pmb{r}+\pmb{P}\pmb{d})+\mbox{
noise terms,}
\end{align}
where the noise terms are data dependent as shown in
\cite{witrisal}.

\section{Joint Demodulation and Equalization Algorithm}
\label{section_decoding_algorithm}

In this section, we present the proposed demodulation and
equalization algorithm. The algorithm is obtained by formulating the
demodulation problem as a nearest neighborhood decoding problem,
reformulating into mixed quadratic programming, and using SDP
relaxation.

In the first step, we formulate the demodulation problem as a
nearest neighborhood decoding problem as follows.
\begin{align}
\label{nearest_neighborhood_decoding} & \min
\sum_{n=0}^{N_b-1}\left\{z[n]-(\pmb{r}+\pmb{P}\pmb{d})^T\pmb{Q}^T\pmb{B}[n]\pmb{Q}(\pmb{r}+\pmb{P}\pmb{d})\right\}^2
\\
& \mbox{subject to } d_n\in\{-1,1\}.
\end{align}
Note that the nearest neighborhood decoding is not the maximal
likelihood decoding in the considered scenario, because noise is
signal dependent.

The above nearest neighborhood decoding problem can be reformulated
as a mixed quadratic programming by introducing auxiliary variables
$s_n$, $n=0,\ldots,N_b-1$.
\begin{align} & \min
\sum_{n=0}^{N_b-1}
\left(s_n\right)^2 \\
& \mbox{subject to } \\
&
s_n=z[n]-(\pmb{r}+\pmb{P}\pmb{d})^T\pmb{Q}^T\pmb{B}[n]\pmb{Q}(\pmb{r}+\pmb{P}\pmb{d}),
\\
&  d_n\in\{-1,1\}.
\end{align}

Now, we claim that the above mixed quadratic programming is
equivalent to the following matrix optimization problem.
\begin{align} & \min
\sum_{n=2}^{N_b+1}
\pmb{U}_{n,n} \label{matrix_opt_eq1}\\
& \mbox{subject to } \\
& \pmb{U}_{1,n}=z[n-2]-\pmb{r}^T\pmb{Q}^T\pmb{B}[n-2]\pmb{Q}\pmb{r}
\nonumber \\
& \hspace{0.5in} -
\pmb{r}^T\pmb{Q}^T\pmb{B}[n-2]\pmb{Q}\pmb{P}\pmb{d}'
\nonumber \\
& \hspace{0.5in} -
\pmb{r}^T\pmb{Q}^T\pmb{B}[n-2]^T\pmb{Q}\pmb{P}\pmb{d}'
\\
& \hspace{0.5in}
-tr\left\{\pmb{D}'\pmb{P}^T\pmb{Q}^T\pmb{B}[n-2]\pmb{Q}\pmb{P}\right\},
\nonumber \\
& \hspace{0.5in} \mbox{for }n=2,\ldots,N_b+1 \nonumber \\
& \pmb{U}_{1,1}=1, \\
 & \pmb{U}_{n,n}=1, \mbox{ for }n=N_b+2,\ldots,2N_b+1,
\\
&  \pmb{d}'=[\pmb{U}_{1,N_b+2},\ldots,\pmb{U}_{1,2N_b+1}]^T, \\
&  \pmb{U}\in \pmb{\mathcal S}, \\
&   \pmb{U}\succeq 0, \\
&  \pmb{U} \mbox{ has rank one}, \label{matrix_opt_eq_final}
\end{align}
where, $\pmb{U}$ denotes a matrix of size $2N_b+1$ by $2N_b+1$,
$\pmb{D}'$ denote the sub-matrix of $\pmb{U}$ formed by selecting
the last $N_b$ rows and columns, and the optimization variables are
the elements of the matrix $\pmb{U}$. Because the matrix $\pmb{U}$
has rank one, is symmetric and positive semi-definite, it is well
known \cite{todd01} that there exists a vector $\pmb{u}$, such that
\begin{align}
\pmb{U}=\pmb{u}\pmb{u}^T.
\end{align}
If we further assume that $\pmb{u}_1=1$, then the vector $\pmb{u}$
is unique. In addition, there is an one-to-one correspondence
between the solution of the mixed quadratic programming and the
solution of the matrix optimization problem,
\begin{align}
\pmb{u}=[1,s_0,s_2,\ldots,s_{N_b-1},d_0,d_1,\ldots,d_{N_b-1}]^T.
\end{align}
Therefore, the mixed quadratic programming is equivalent to the
matrix optimization problem.

In the matrix optimization problem, all objective function and
constraints are convex except the rank one constraint in Eq.
\ref{matrix_opt_eq_final}. If the rank one constraint is relaxed,
then we obtain  a convex relaxation (SDP relaxation). The convex
optimization problem can then be efficiently solved by
polynomial-time algorithms and softwares, for example, by using the
SeDuMi package \cite{sturm}. Previous research has shown that such
SDP relaxations are tight approximations to the original problems
\cite{goemans}. Near-optimal solutions of the original problems can
be obtained from SDP relaxations by random rounding.

Finally, the proposed joint demodulation and equalization algorithm
consists of two steps. In the first step,  the following SDP
relaxation problem is solved.
\begin{align} & \min
\sum_{n=2}^{N_b+1}
\pmb{U}_{n,n} \label{sdp_opt_eq1}\\
& \mbox{subject to } \\
& \pmb{U}_{1,n}=z[n-2]-\pmb{r}^T\pmb{Q}^T\pmb{B}[n-2]\pmb{Q}\pmb{r}
\nonumber \\
& \hspace{0.5in} -
\pmb{r}^T\pmb{Q}^T\pmb{B}[n-2]\pmb{Q}\pmb{P}\pmb{d}'
\nonumber \\
& \hspace{0.5in} -
\pmb{r}^T\pmb{Q}^T\pmb{B}[n-2]^T\pmb{Q}\pmb{P}\pmb{d}'
\\
& \hspace{0.5in}
-tr\left\{\pmb{D}'\pmb{P}^T\pmb{Q}^T\pmb{B}[n-2]\pmb{Q}\pmb{P}\right\},
\nonumber \\
& \hspace{0.5in} \mbox{for }n=2,\ldots,N_b+1 \nonumber \\
& \pmb{U}_{1,1}=1, \\
 & \pmb{U}_{n,n}=1, \mbox{ for }n=N_b+2,\ldots,2N_b+1,
\\
&  \pmb{d}'=[\pmb{U}_{1,N_b+2},\ldots,\pmb{U}_{1,2N_b+1}]^T, \\
&  \pmb{U}\in \pmb{\mathcal S}, \\
&   \pmb{U}\succeq 0, \\
& \pmb{D}'\mbox{ is the submatrix of }\pmb{U}
 \mbox{ formed by selecting the last }N_b\nonumber \\
&\mbox{rows and columns}.
\end{align}
In the second step, the demodulation decision is made by
thresholding,
\begin{align}
d_n=\mbox{sign}(\pmb{U}_{1,n+N_b+2}).
\end{align}

\section{Numerical Results }

\label{sec_numerical}

In this section, we present simulation results for the proposed
demodulation and equalization scheme. We assume that the channel can
be modeled by the S-V model \cite{saleh87}. The received signal for
each transmitted pulse $\bar{w}(t)$ is,
\begin{align}
g(t)=\sum_{j=1}^{N_m}\alpha_jw(t-\delta_j).
\end{align}
In the above equation, $w(t)$ is the second derivative Gaussian
monocycle,
\begin{align}
w(t)=\left[1-4\pi\left(t/\tau_m\right)^2\right] \exp\left\{-2\pi
\left(t/\tau_m\right)^2\right\}
\end{align}
where $\tau_m=0.2877$ nanosecond. $N_m$ is the total number of
multiple paths. $\alpha_j$ and $\delta_j$ are amplitude and delay of
the $j$-th path.

We assume that the delays of the paths follow the Poisson process
with the expected interval between two consecutive paths being $10$
nanoseconds. The amplitude is Raleigh distributed, such that the
expectation of the amplitude $\alpha_j$ is $\exp(-\delta_j/T_e)$,
where $T_e=20$ nanoseconds.  We assume that the amplitudes and
delays $\alpha_j$, $\delta_j$ vary slowly, so that the matrices
$\pmb{B}[n]$ can be accurately estimated (for example, by using
pilot signals), and considered perfectly known at the demodulator.

\comment{
\begin{figure}[h]
 \centering
 \includegraphics[width=3in]{delay_spread.eps}
 \caption{A realization of $\alpha_j$ and $\delta_j$}
 \label{delay_spread}
\end{figure}}

For the transmitted signal, we assume that each block has $N_b=10$
symbols and each symbol corresponds to $N_p=4$ pulses. The
pseudo-random delay hopping code is $[1.7,1.9,2.1,2.3]$. The symbol
duration $T_s=8$ nanoseconds. The integral time $T_I=T_s=8$
nanoseconds.

\begin{figure}[h]
 \centering
 \includegraphics[width=3.5in]{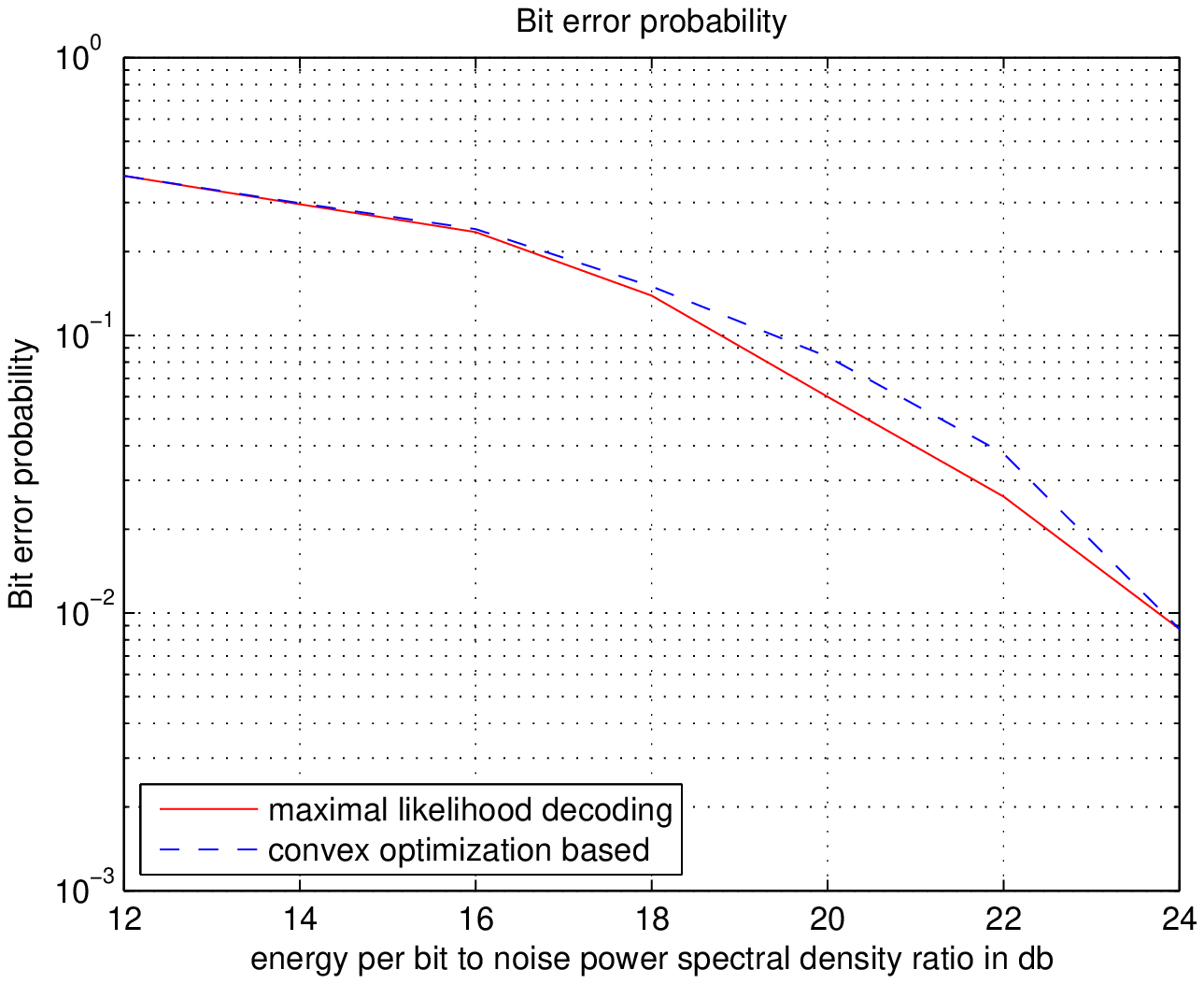}
 \caption{Bit error probabilities  in the case of severe non-linearity. The length of delay
 spread is $200$ nanoseconds. The solid curve represents the bit
 error probabilities of the maximal likelihood detection algorithm.
 The dashed curve represents the bit error probabilities of the
 proposed reduced complexity detection algorithm. The X-axis shows
 energy per bit to noise power spectral density ratio $E_b/N_0$ in
 dB.}
 \label{200ns_bit}
\end{figure}

We illustrate the bit error probability of the proposed demodulation
and equalization algorithm in three different cases. In the first
case, we assume that the delay spread extends over a range of $200$
nanoseconds. Therefore, there exists severe non-linearity in the
system. The bit error probability of the proposed scheme for this
case is shown in Fig. \ref{200ns_bit}. The bit error probability of
the maximal likelihood detection algorithm is also plotted.

\begin{figure}[h]
 \centering
 \includegraphics[width=3.5in]{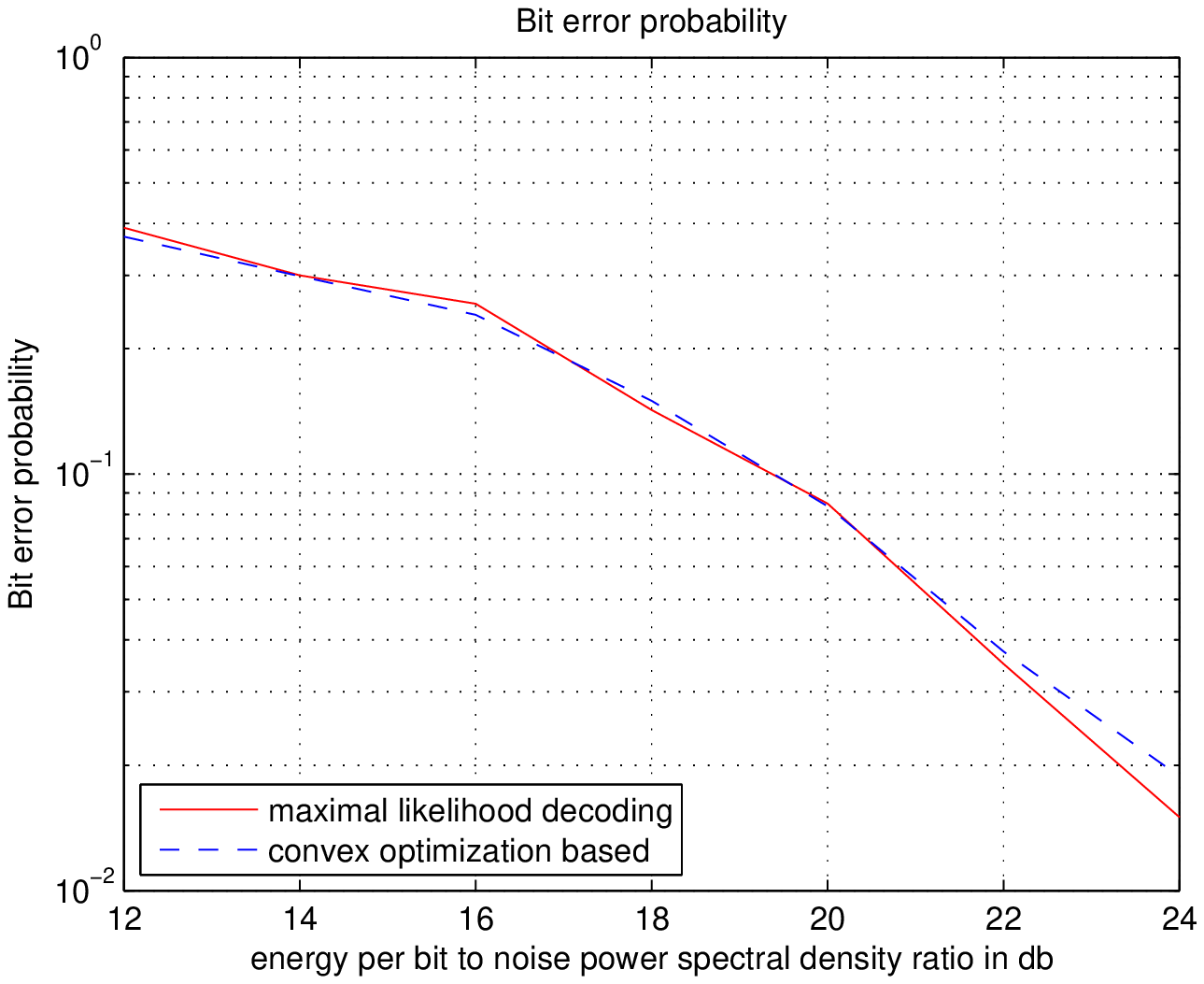}
 \caption{Bit error probabilities  in the case of mild non-linearity. The
 length of delay
 spread is $30$ nanoseconds. The solid curve represents the bit
 error probabilities of the maximal likelihood detection algorithm.
 The dashed curve represents the bit error probabilities of the
 proposed reduced complexity detection algorithm. The X-axis shows
 energy per bit to noise power spectral density ratio $E_b/N_0$ in
 dB.}
 \label{30ns_bit}
\end{figure}

\begin{figure}[h]
 \centering
 \includegraphics[width=3.5in]{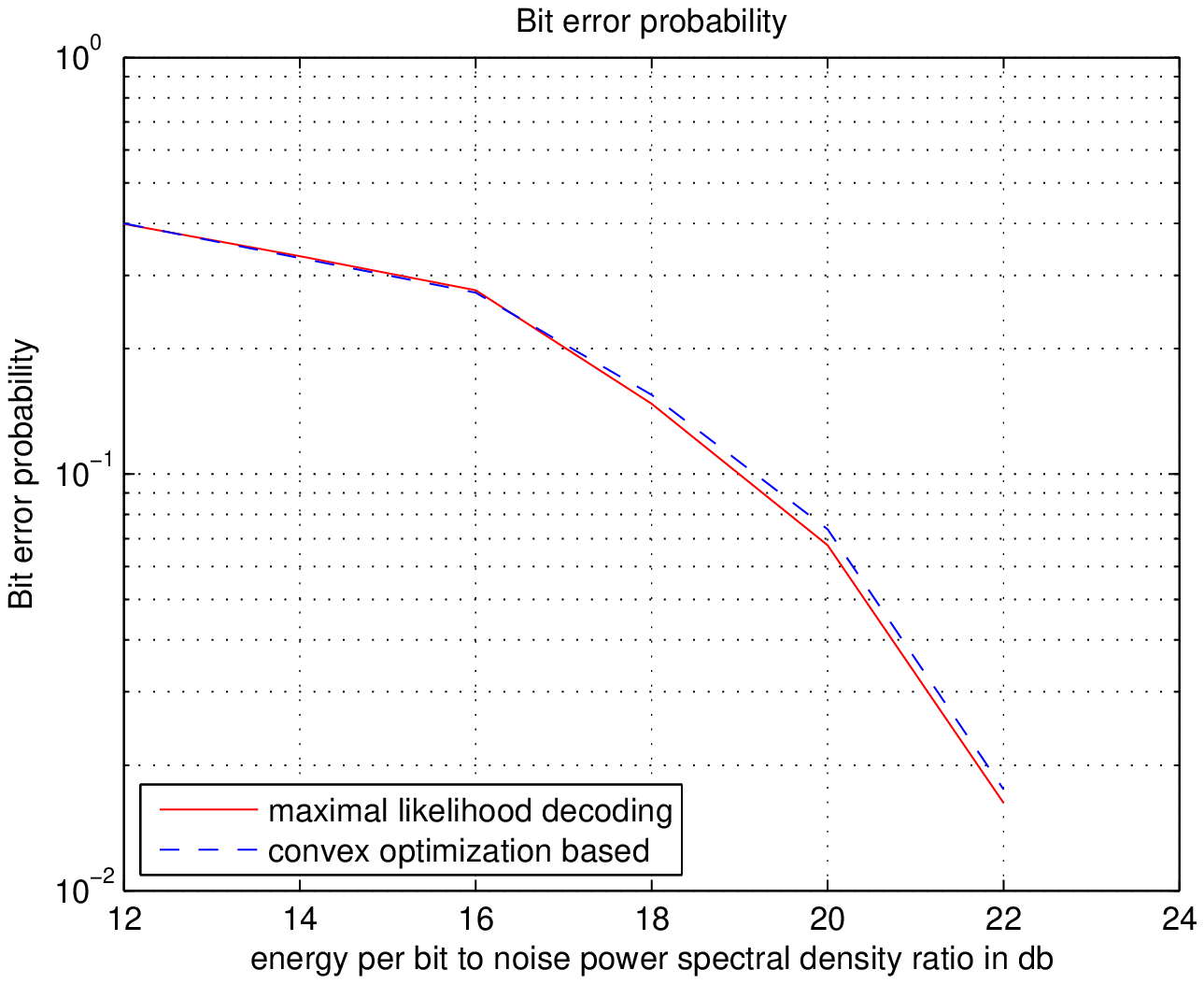}
 \caption{Bit error probabilities  in the case of linear channels.
 The frequency response of the channel is flat. The solid curve represents the bit
 error probabilities of the maximal likelihood detection algorithm.
 The dashed curve represents the bit error probabilities of the
 proposed reduced complexity detection algorithm. The X-axis shows
 energy per bit to noise power spectral density ratio $E_b/N_0$ in
 dB.
 }
 \label{0ns_bit}
\end{figure}

In the second case, we assume that the delay spread extends over a
range of $30$ nanoseconds. The non-linearity is mild in this case.
The bit error probabilities of the proposed scheme and the maximal
likelihood decoding are shown in Fig. \ref{30ns_bit}. In the third
case, we assume that there is no ISI. And the bit error
probabilities are shown in Fig. \ref{0ns_bit}. From all these
results, we conclude that even though the proposed scheme is
sub-optimal, the performance loss is negligible.

\comment{The pseudo-random amplitude code is [-1,-1,1,1]}

\section{Conclusion}

\label{sec_conclusion}

In this paper, we propose a convex optimization based demodulation
and equalization algorithm with low complexity for the differential
IR UWB systems. The complexity of the proposed algorithm grows
polynomially with respect to the blocklengths, and is independent of
the length of delay spread. Even though the proposed algorithm is
sub-optimal, we show by simulation results that the performance loss
is negligible. The proposed demodulation and equalization algorithm
is a near-optimal algorithm with significantly reduced computational
complexity.

\bibliographystyle{IEEEtran}
\bibliography{the_bib}

\end{document}